\documentclass[3p,times,procedia]{elsarticle}
\usepackage{nupha_ecrc}

\volume{00}

\firstpage{1}

\journalname{Nuclear Physics A}

\runauth{H. Petersen et al.}

\jid{nupha}

\jnltitlelogo{Nuclear Physics A}

\usepackage{amssymb}

\usepackage[figuresright]{rotating}

\begin{document}

\begin{frontmatter}



\dochead{}

\title{Influence of kinematic cuts on the net charge distribution}


\author[label1,label2,label3]{Hannah Petersen}
\author[label1,label4]{Dmytro Oliinychenko}
\author[label1]{Jan Steinheimer}
\author[label1,label2]{Marcus Bleicher}

\address[label1]{Frankfurt Institute for Advanced Studies, Ruth-Moufang-Str. 1, 60438 Frankfurt am Main, Germany}
\address[label2]{Institut f\"ur Theoretische Physik, Goethe Universit\"at, Max-von-Laue-Str. 1, 60438 Frankfurt am Main, Germany}
\address[label3]{GSI Helmholtzzentrum f\"ur Schwerionenforschung GmbH, Planckstr. 1, 64291 Darmstadt, Germany} 
\address[label4]{Bogolyubov Institute for Theoretical Physics, Kiev 03680, Ukraine}

\begin{abstract}
The higher moments of the net charge distributions, e.g. the skewness and kurtosis, are studied within an infinite hadronic matter calculation in a transport approach. By dividing the box into several parts, the volume dependence of the fluctuations is investigated. After confirming that the initial distributions follow the expectations from a binomial distribution, the influence of quantum number conservation in this case the net charge in the system on the higher moments is evaluated. For this purpose, the composition of the hadron gas is adjusted and only pions and $\rho$ mesons are simulated to investigate the charge conservation effect. In addition, the effect of imposing kinematic cuts in momentum space is analysed. The role of resonance excitations and decays on the higher moments can also be studied within this model. This work is highly relevant to understand the experimental measurements of higher moments obtained in the RHIC beam energy scan and their comparison to lattice results and other theoretical calculations assuming infinite matter. 
\end{abstract}


\end{frontmatter}


\section{Introduction}
\label{intro}



One of the main goals of heavy ion research is the exploration of the phase diagram of strongly interacting matter. At high temperatures and low baryo-chemical potentials lattice QCD calculations have established a cross-over transition between the hadron gas and the quark gluon plasma phase. It is expected that at lower temperatures and higher densities the transition turns into a first order phase transition separated by a critical endpoint from the cross-over region. Since there are no first principle calculations available up to date it is crucial to establish experimental observables that indicate the critical behaviour of the system formed in heavy ion collisions. Higher moments of the net charge and net baryon number distributions, e.g. the skewness and kurtosis, are very promising for this purpose, since they are sensitive to higher powers of the correlation length \cite{Asakawa:2015ybt}. If strongly interacting matter passes through a critical endpoint the correlation length increases and this is imprinted in the higher moments of the particle distributions \cite{Berdnikov:1999ph,Skokov:2012ds}. Since the quark gluon plasma is formed in heavy ion collisions as a transient state of finite volume, it is important to investigate the dynamical evolution of the system when passing through the critical region \cite{Nahrgang:2011mg}. The STAR collaboration has recently measured the beam energy dependence of higher moments \cite{Adamczyk:2014fia}.  The measurements of multiplicity or conserved charge fluctuations are affected by a multitude of other effects such as volume fluctuations due to centrality cuts, global conservation laws, detector efficiencies and finite statistics \cite{Luo:2013bmi, Bzdak:2013pha, Bzdak:2012ab, Bzdak:2012an, Karsch:2015zna} that have to be taken into account before drawing conclusions about the phase diagram. 
In this work, the effect of cuts in momentum space on the higher moments of the net charge distribution is studied within a hadronic transport approach.

\section{Higher moments in the transport approach}
\label{model}
\begin{figure}
\includegraphics[width=0.5\textwidth]{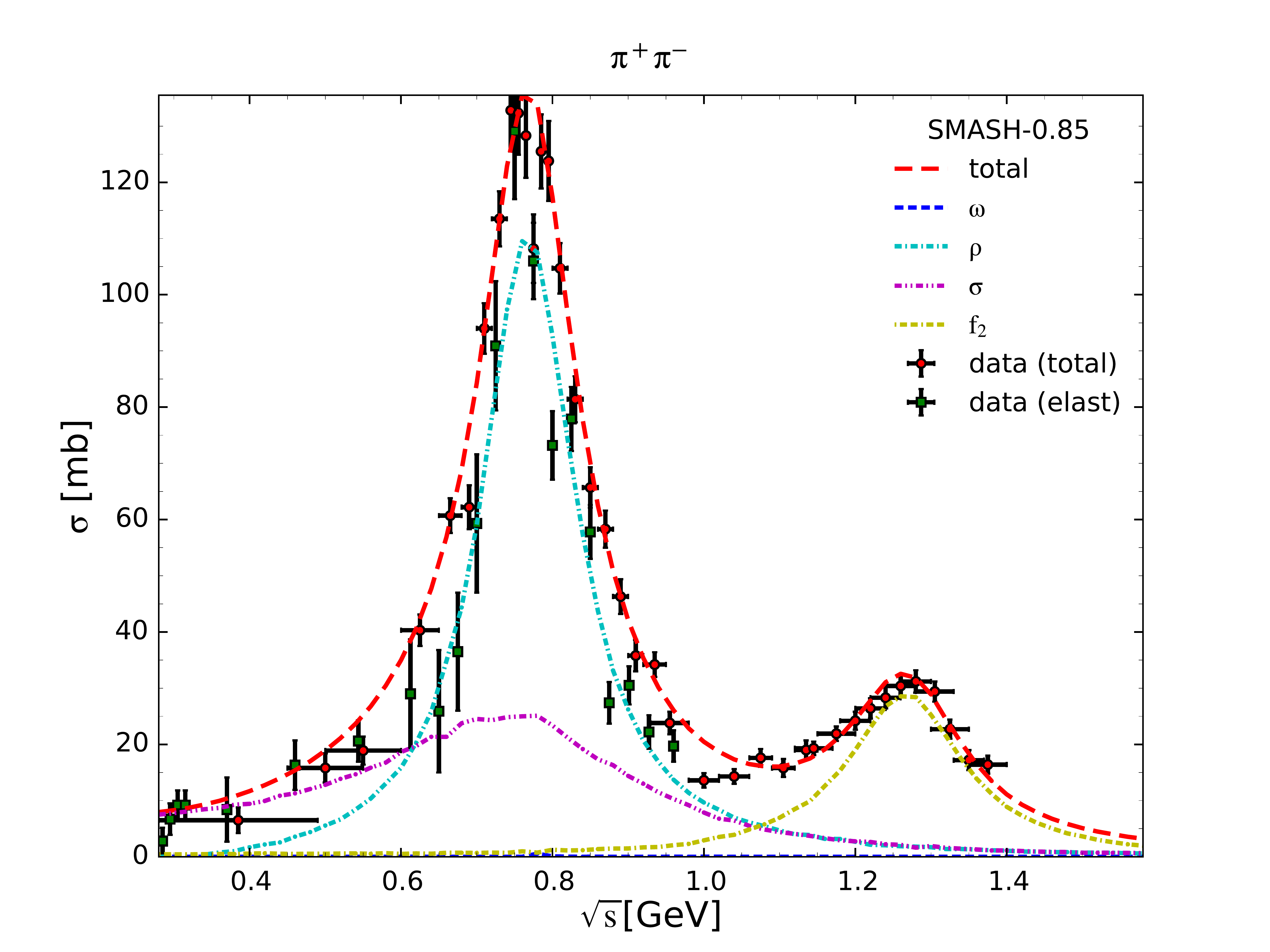}
\includegraphics[width=0.5\textwidth]{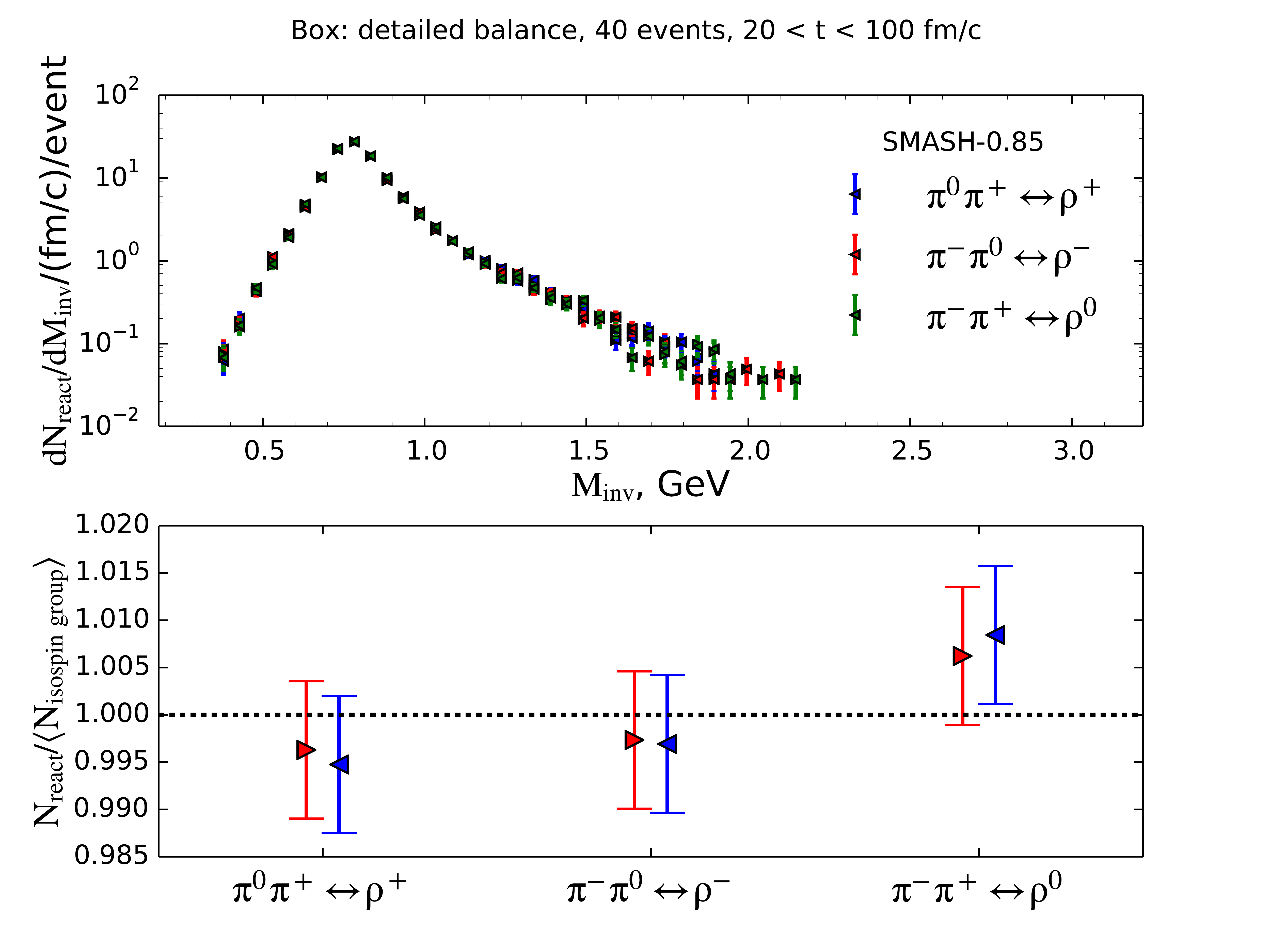}
\caption{Left: The different channels contributing to the $\pi^+-\pi^-$ cross-sections are shown including a comparison to the PDG values \cite{Beringer:1900zz} for the total and elastic cross section. Right: Detailed balance over the whole invariant mass distribution (upper panel) and summarized for individual reactions (lower panel).
\label{fig_pi_pi_xs}}
\end{figure}

Infinite hadronic matter is calculated within a hadronic transport approach SMASH (Simulating Many Accelerated Strongly-interacting Hadrons) in a box (V=$10^3$ fm$^3$) with periodic boundary conditions. There is either one particle species with purely elastic scattering or pions that are allowed to form $\rho$ mesons. Fig. \ref{fig_pi_pi_xs} (left) shows the (in) elastic cross-section of pions as a function of the center of mass energy of the binary collision in comparison to experimental data. Detailed balance is fulfilled in all reactions and in all phase space bins as shown in Fig. \ref{fig_pi_pi_xs} (right). On the order of 5 million events are simulated to reduce the statistical error bars for the kurtosis calculation. The volume of the box is divided into 10 intervals ($\Delta x = 1$ fm) and the higher moments are calculated from the final particles after resonance decays. We have checked that the initial distribution follows the expectations from a binomial distribution and free streaming  does not change this result. The statistical erros for the skewness and kurtosis are estimated following the procedure explained in \cite{Luo:2014rea}.

\begin{figure}
\centering
\includegraphics[width=0.4\textwidth]{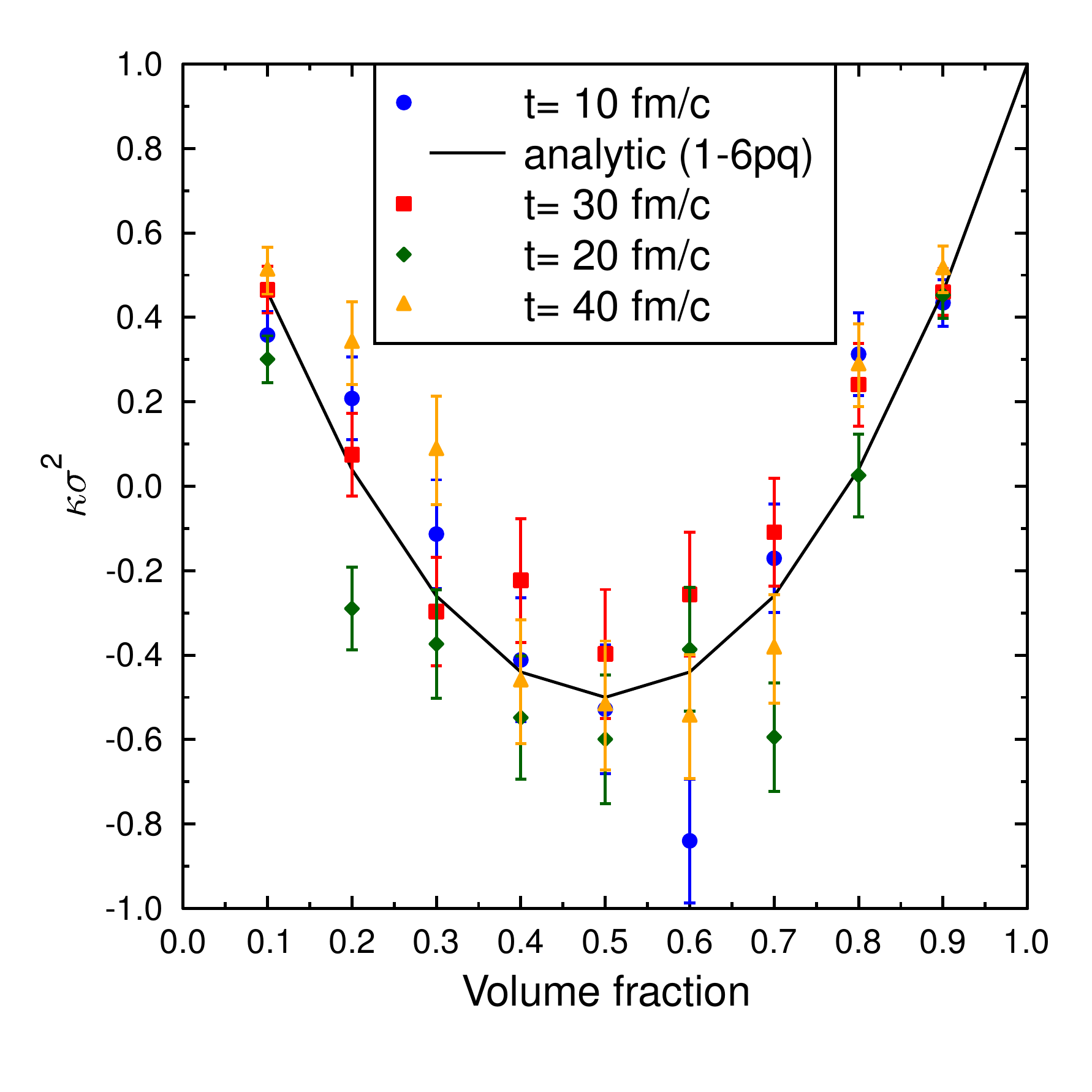}
\includegraphics[width=0.4\textwidth]{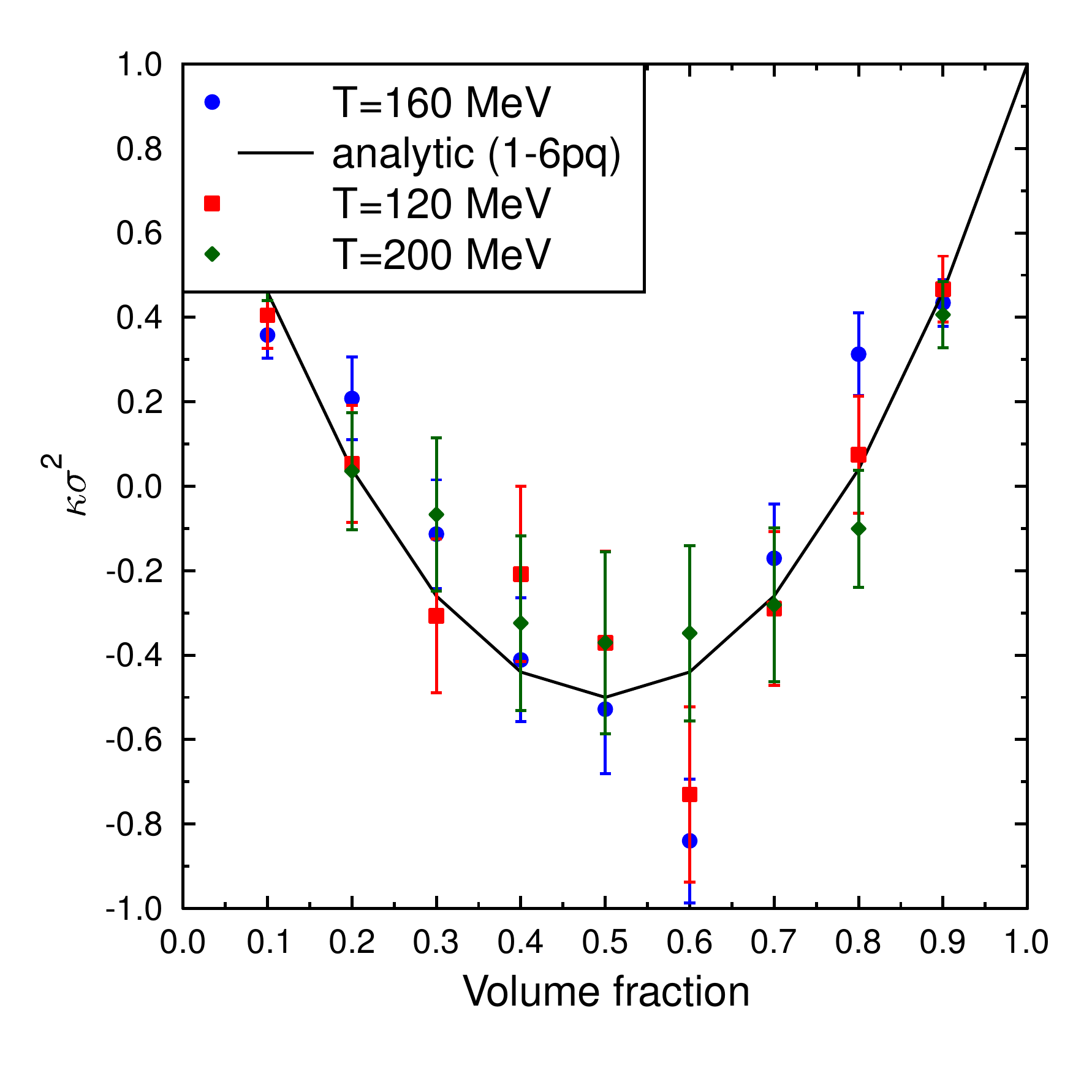}
\caption{Left: Time evolution of the kurtosis in a box with 400 pions with only elastic scatterings ($\sigma=10$ mb). Right: Temperature dependence of the kurtosis in a box with 400 pions and only elastic scatterings ($\sigma=10$ mb).
\label{fig_time_evolution}}
\end{figure}

In Fig. \ref{fig_time_evolution} (left), it is shown, that the kurtosis saturates after 10 fm/c time evolution to the values expected for a binomial distribution and does not change systematically afterwards. Therefore, all calculations are done after $t=10$ fm/c. In addition, for a box with 400 pions and an elastic cross-section $\sigma=10$ mb, there is no temperature dependence of the results for the kurtosis in the range from $T=120-200$ MeV (see Fig. \ref{fig_time_evolution}, right). Therefore, we use a temperature of $T=160$ MeV in the following.  

\section{Results for the $\pi-\rho$ system}
\label{results}

\begin{figure}
\centering
\includegraphics[width=0.4\textwidth]{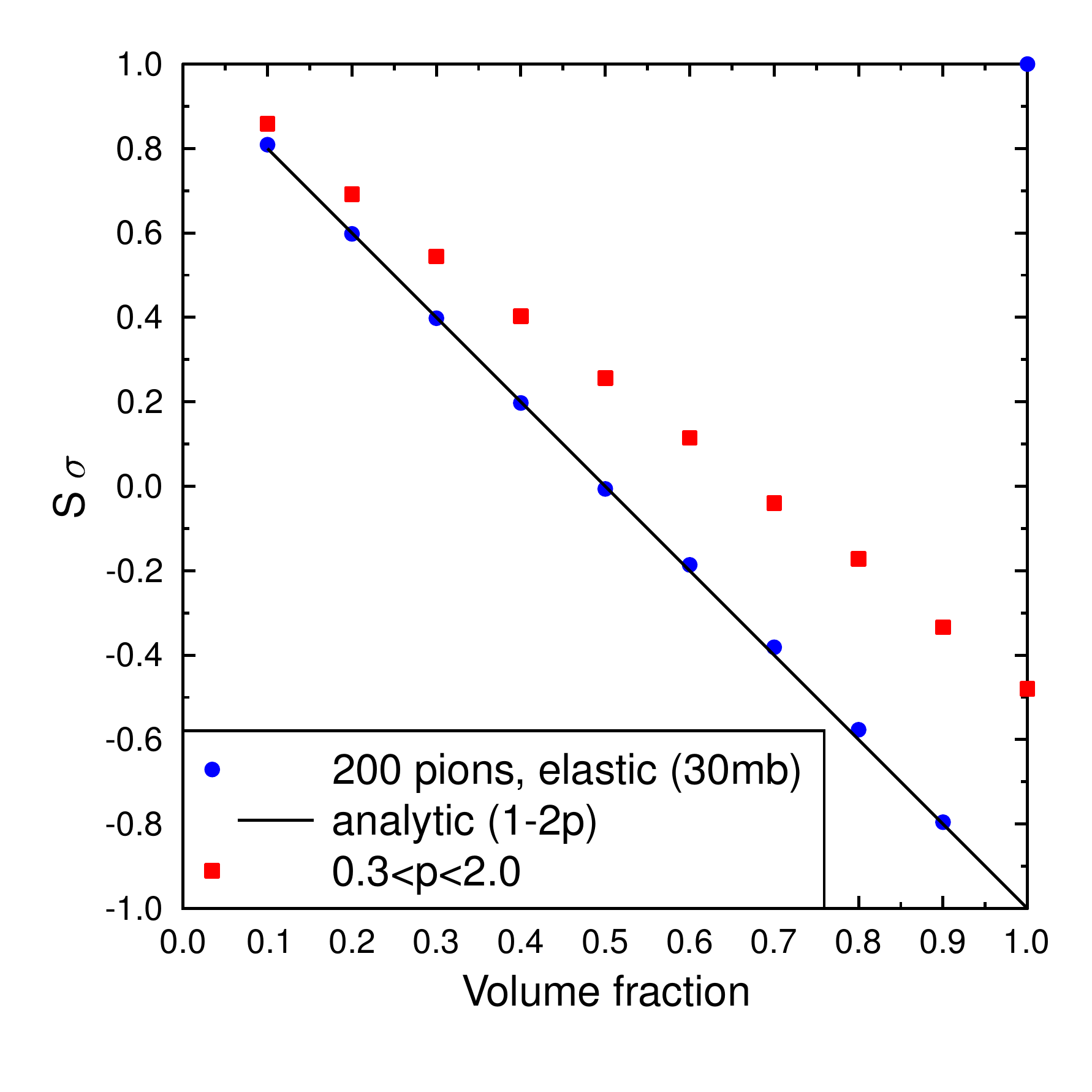}
\includegraphics[width=0.4\textwidth]{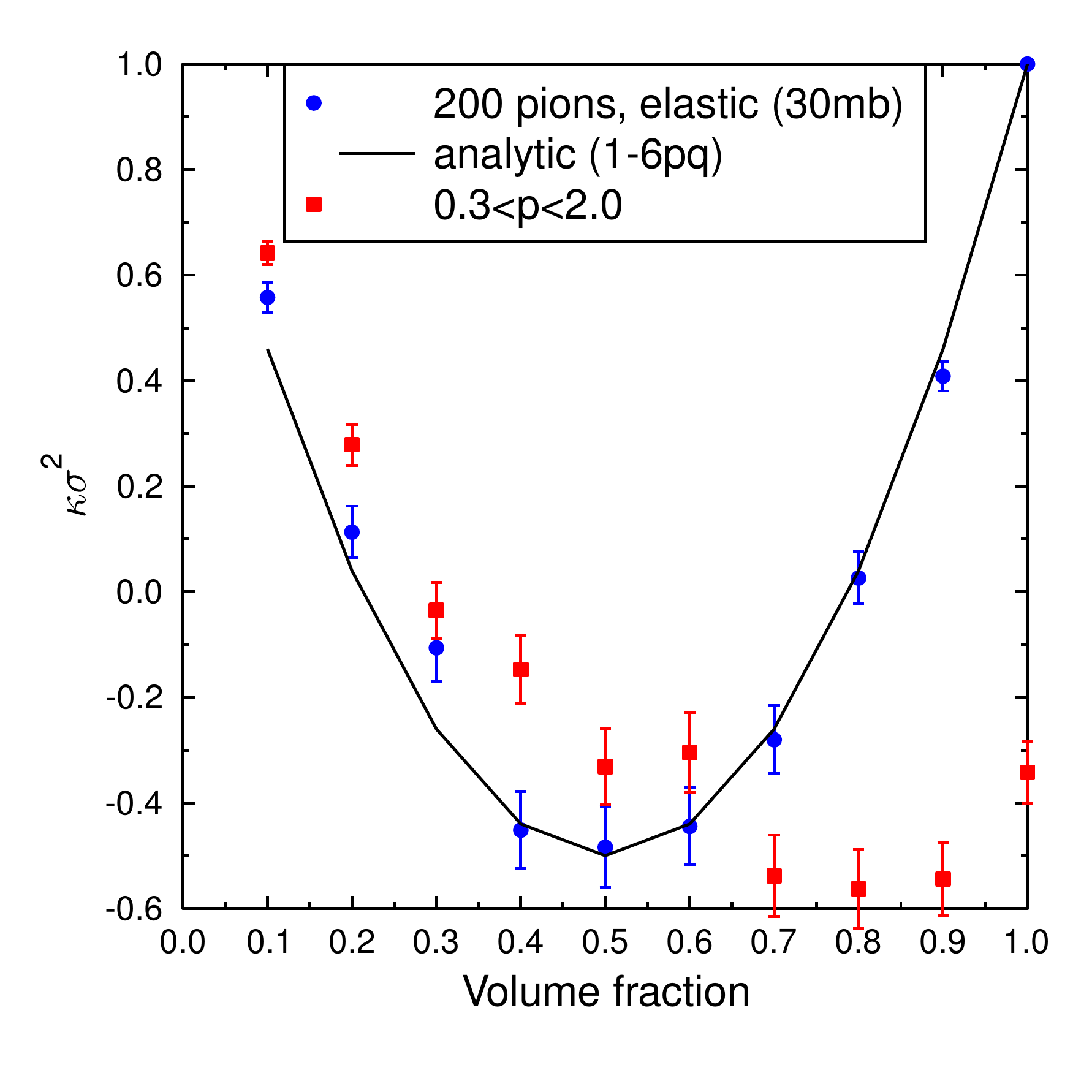}
\caption{Influence of a momentum cut on the skewness (left) and kurtosis (right) in a box with 200 pions and elastic cross section $\sigma=30$ mb. 
\label{fig_mom_cut_skew}}
\end{figure}

Before coming to the more interesting $\pi-\rho$ system, let us have a look at the influence of momentum space cuts in a box with 200 pions and a constant elastic cross-section of $\sigma=30$ mb. In experiments kinematic cuts are required for acceptance and efficiency reasons, therefore we choose a realistic cut of $0.3<p<2$ GeV where $p$ is the absolute value of the 3-momentum. In the case of a heavy ion reaction that would correspond to a cut in transverse momentum. In Fig. \ref{fig_mom_cut_skew} the skewness and kurtosis without momentum cuts agree well with the analytic expectation for a binomial distribution. In case of a momentum cut, the full volume does not contain the total number of particles anymore, therefore the moments are "shifted" to the right, but still show the same trends as before. 

\begin{figure}
\centering
\includegraphics[width=0.4\textwidth]{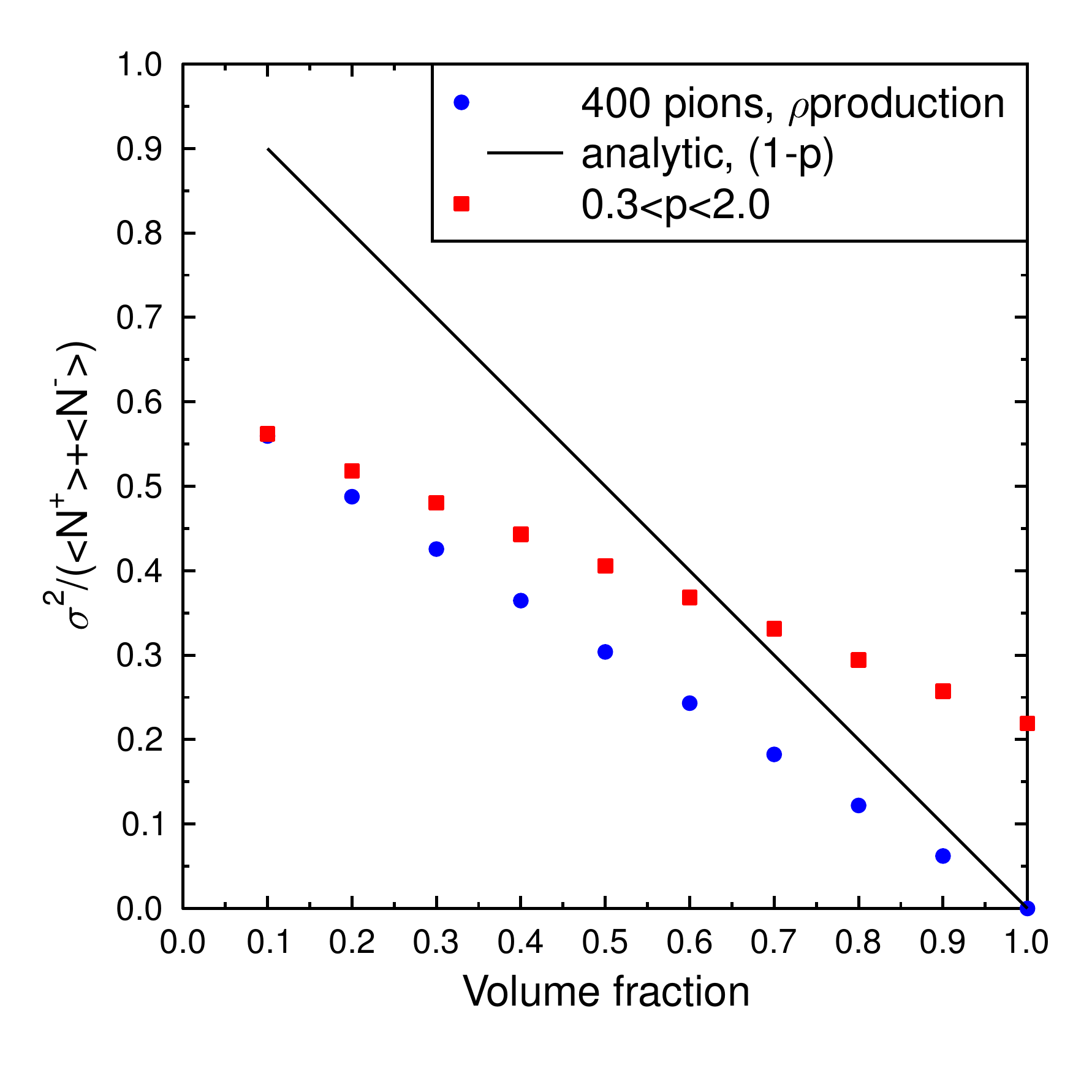}
\includegraphics[width=0.4\textwidth]{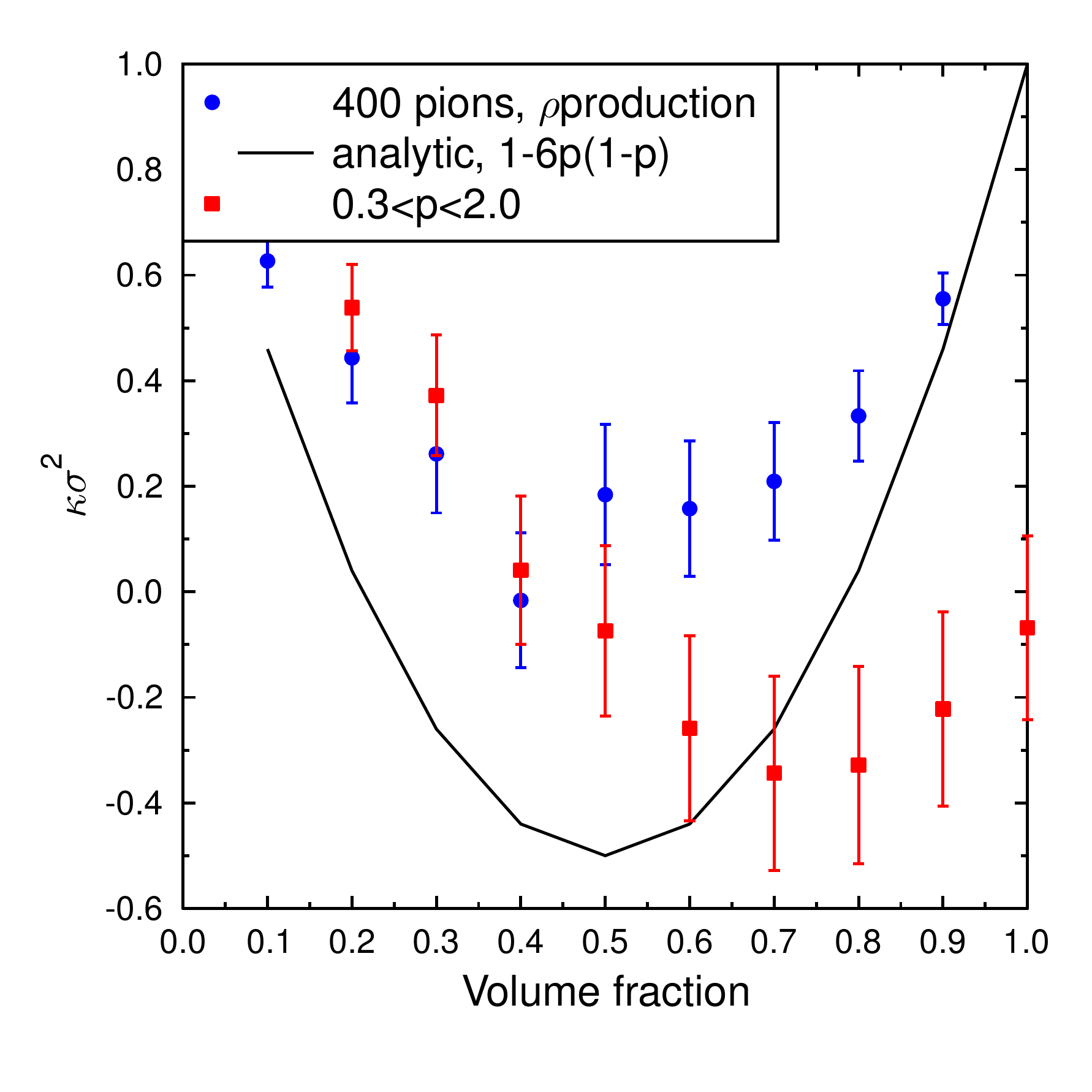}
\caption{Influence of local charge conservation and momentum cut on the variance (left) and kurtosis (right) of the net charge dsitribution in a box with 200 $\pi^+$ and 200 $\pi^-$ with $\rho$ production.
\label{fig_net_moments}}
\end{figure}

In the box with 200 $\pi^+$ and 200 $\pi^-$ each of the charges separately is distributed according to a binomial distribution. Therefore, the net charge of the system follows a Skellam distribution with mean equals to zero. Therefore, the skewness is also equal to zero and we concentrate on the variance and kurtosis instead. In Fig. \ref{fig_net_moments} two counteracting effects can be observed. The production and decay of $\rho$ mesons introduces a local charge conservation (blue circles) and therefore suppresses the fluctuations compared to the Skellam expectation. The momentum cut on the other hand increases the flucuations, since the total particle number in the volume is not conserved anymore (red squares). 

\section{Conclusion and Outlook}
\label{concl}
In this study a hadronic transport approach has been employed to investigate the effects of resonance formation and kinematic cuts on the higher moments of the particle distributions. First, simple tests show, that the initialization fulfills the binomial expectation as well as there is no time and temperature dependence of the results. Local charge conservation due to resonance formation suppresses the fluctuations whereas momentum cuts increase the fluctuations, since the total particle number in the full volume is not conserved anymore. 
In the future, this approach will be used to study more complicated systems involving baryonic and mesonic degrees of freedom to study the effect of baryon diffusion, global and local conservation laws and kinematic cuts on the higher moments of the net baryon, net proton and net charge distribution.

\section*{Acknowledgements}
This work was supported by the Helmholtz International Center for the Facility for Antiproton and Ion Research (HIC for FAIR) within the framework of the Landes-Offensive zur Entwicklung Wissenschaftlich-Oekonomischer Exzellenz (LOEWE) program launched by the State of Hesse. HP acknowledges funding of a Helmholtz Young Investigator Group VH-NG-822 from the Helmholtz Association and GSI. DO acknowledges support of the Deutsche Telekom Stiftung. Computational resources have been provided by the Center for Scientific Computing (CSC) at the Goethe University of Frankfurt. 





\end{document}